\documentstyle{article}

\newcommand{\ba}{\begin{eqnarray}}
\newcommand{\ea}{\end{eqnarray}}
\newcommand{\ii}{\'\i}
\begin{document}
\pagestyle{plain}

\title{A Symmetry-Adapted Algebraic 
 Approach to Molecular Spectroscopy}
\author{A. Frank$^{1,2)}$, R. Lemus$^{1)}$, R. Bijker$^{1)}$, 
F. P\'erez-Bernal$^{3)}$ and J.M. Arias$^{3)}$\\
\and
\begin{tabular}{rl}
$^{1)}$ & Instituto de Ciencias Nucleares, U.N.A.M.,\\
        & A.P. 70-543, 04510 M\'exico D.F., M\'exico\\
$^{2)}$ & Instituto de F\'{\i}sica, Laboratorio de Cuernavaca,\\
        & A.P. 139-B, Cuernavaca, Morelos, M\'exico\\
$^{3)}$ & Departamento de F\'{\i}sica At\'omica, Molecular y Nuclear,\\
        & Facultad de F\'{\i}sica, Universidad de Sevilla,\\
        & Apdo. 1065, 41080 Sevilla, Espa\~na
\end{tabular}}
\date{}
\maketitle

\section{Introduction}

The study of molecular vibrational spectra \cite{uno} requires
theoretical models in order to analyze and interpret the
measurements.  These models range from simple
parametrizations of the energy levels, such as the Dunham
expansion \cite{dos}, to {\it ab initio} calculations, where solutions
of the Schr\"odinger equation in different approximations are
sought \cite{tres,cuatro,cinco,veintinueve}.  
In general, the latter involve the
use of internal coordinates and the evaluation of force field
constants associated to derivatives at the potential minima.  While
this method can be reliably applied to small molecules \cite{seis}, it
quickly becomes a formidable problem in the case of larger
molecules, due to the size of their configuration spaces.  New
calculational tools to describe complex molecules are thus needed. 

In 1981 an  algebraic approach was 
proposed to describe the roto-vibrational structure of diatomic
molecules \cite{siete},   subsequently  extended to  linear
tri- and 
four- atomic molecules \cite{ocho} and certain  non-linear  triatomic
molecules \cite{nueve}.  Although these   were encouraging results, the
model  could not   
be extended to  polyatomic molecules,  due to the impossibility of 
 incorporating  the  underlying 
discrete symmetries.  This difficulty could be surmounted by treating
the vibrational degrees of freedom separately from the rotations.  In
1984 Van Roosmalen {\it et al.} proposed  a $U(2)$ based model
to describe the stretching vibrational modes in ABA
molecules \cite{diez} which was later 
extended to describe the stretching vibrations of polyatomic
molecules such as octahedral and benzene like molecules \cite{once}. 
Recently the bending modes have also been included in  the
framework, which was subsequently  applied to 
 describe ${\cal
C}_{2v}$-triatomic molecules \cite{doce} and the lower excitations of
tetrahedral molecules \cite{trece}, using  
a scheme which combines Lie-algebraic and point group methods.  
 In a different approach, it has also been 
 suggested \cite{catorce} to use a $U(k+1)$ model for the $k = 3n-3$ 
rotational and vibrational degrees of freedom of a $n$-atomic
molecule.  This model has the advantage that it  incorporates
all rotations and vibrations and takes into  account
 the relevant point group symmetry, but for larger
molecules the  number of possible interactions and the 
size of the Hamiltonian matrices increase very
rapidly, making  it impractical to apply. 

 Although
 the  algebraic formulations have  proved useful,  several
problems remained, most important of which is the 
absence of a clear connection to traditional methods.  On the other hand, a
related problem  is the lack of a systematic
procedure to construct all physically meaningful interactions in the
algebraic space.  In this paper we show that both these issues can be
resolved  by considering a symmetry-adapted version of the $U(2)$
algebraic model  for the analysis of molecular vibrational 
spectra. In this approach  it is possible to construct algebraic
operators with  well defined physical meaning,   in
particular  interactions  fundamental 
 for the description of the degenerate modes
present in systems exhibiting high degree of symmetry.  The
 procedure to construct them   takes full advantage of
the discrete  symmetry of the molecule and  gives rise to  all
possible terms in a 
systematic fashion,  providing  a clear-cut connection   
 between the algebraic scheme and the traditional
analyses based on internal coordinates, which  correspond to   the
harmonic limit of the model \cite{diecisiete}. 

As a test of the symmetry-adapted approach 
we discuss an application to three 
${\cal D}_{3h}$-triatomic molecular systems, namely H$^+_3$, Be$_3$ and
Na$^+_3$, and to two tetrahedral molecules, the Be$_4$ cluster and
the methane molecule.   Since  small molecules can in general 
be      well described by means of {\it ab initio}
calculations \cite{quince,dieciseis}, we  emphasize the basic purpose 
 of this work.
We establish  an exact  correspondence
between configuration space and algebraic interactions by studying
the harmonic limit of the $U(2)$ algebra.  
This general procedure not only allows to derive algebraic 
interactions from  interactions in 
configuration space, but can also  be applied to cases for which no
configuration space interactions are available.  
  The ${\cal D}_{3h}$-triatomic molecules constitute the
simplest systems where degenerate modes appear and where the new
interactions in the  model become significant. In the case of Be$_4$
we present 
a  comparison with {\it ab initio} calculations, while for CH$_4$ we
present a detailed comparison with experiment.  
  The application 
of these techniques to  other molecules, as well as a more complete
presentation can be found in references 
\cite{diecisiete,X3,ozono,metano}.   

\section{The U(2) vibron model}

The   model  is based on  the isomorphism of the $U(2)$ Lie
algebra and the one dimensional Morse oscillator 
\ba
{\cal H} = - {\hbar^2 \over 2\mu} {d^2\over dx^2} + D (e^{-2x/d} 
- 2e^{-x/d} ) ~~ ,  
\ea
whose eigenstates ${\cal E}$ can be associated with  $U(2)\supset
 SO(2)$ states \cite{dieciocho}.  In order to see how this isomorphism comes
about, consider the radial equation 
\ba
{1\over 2} \biggl( - {1\over r} {d\over dr} r {d\over dr} + {\sigma^2
\over r^2} + r^2 \biggr) \phi (r) = (N+1) \phi (r) ~~ , 
\ea
which corresponds to a two-dimensional harmonic oscillator (in units where
$\hbar = \mu = e = 1)$ associated to a $U(2)$ symmetry
algebra \cite{diecinueve}.  By carrying out a change of variable
$$
r^2 = (N+1)e^{-\rho} ~~ , 
$$
Eq.~(2) transforms into 
\ba
\biggl[ - {d^2 \over d\rho^2} + \biggl( {N + 1 \over 2}\biggr)^2
(e^{-2\rho} - 2e^{-\rho}) \biggr] \phi(\rho) 
= - \biggl( \frac{\sigma}{2} \biggr)^2 \phi (\rho) ~~ .
\ea
This can be identified with Eq.~(1) after defining $x = \rho d$
and  multiplying by $\hbar^2/2\mu d^2$, provided that 
\ba 
D  & = & { \hbar^2 \over 8\mu d^2} (N+1)^2 ~~ ,   \\
{\cal E} & = & - {\hbar^2 \over 2\mu d^2} m^2 ~~ ,
\ea
where we have defined $m=\sigma/2$.
In the framework of the $U(2)$ algebra, the operator  $\hat N$
corresponds to the total 
number of bosons and is fixed by the potential shape according to
Eq.~(4), while $m$, the eigenvalue of the $SO(2)$ generator $J_z$,
takes the values $m = \pm N/2$, $\pm (N-2)/2, \dots$~.  The Morse
spectrum is reproduced twice and consequently for these applications
the $m$-values must be restricted to be positive.  In terms of the
  $U(2)$ algebra, it is  clear from Eqs.~(3-5) that the Morse
Hamiltonian has the algebraic realization 
\ba
\hat  H = - {\hbar^2 \over 2\mu d^2} \hat J^2_z = - A \hat J^2_z ~~.
\ea
In addition, the $U(2)$ algebra includes the raising and lowering
operators $\hat J_+$ and $\hat J_-$, which connect different energy
states, while the angular momentum operator is given by
$\hat J^2 = \hat N ( \hat N+2)/4$, as can be readily shown. 

The Morse Hamiltonian of Eq.~(6) can be rewritten in the more convenient
form  
\ba
\hat H^{\prime} = \hat H + A {\hat N^2 \over 4} =
{A \over 2} [ ( \hat J_ + \hat J_- + \hat J_-
\hat J_+) - \hat N ] ~~ ,  
\ea
where we have used the relation $\hat J^2_z = \hat J^2 - 
(\hat J_+ \hat J_- + \hat J_- \hat J_+)/2$ and  added a constant
term $A \hat N^2/4$ in order to place the ground state
at zero energy.  The parameters $N$ and $A$ are related to the usual  
 harmonic and anharmonic constants $\omega_e$ and
$x_e\omega_e$ used in spectroscopy. To obtain this
relation it is convenient to introduce the quantum number 
\ba
v = {N \over 2} - m ~~ , 
\ea
which corresponds to the number of quanta in the
oscillator. In terms of $v$, the corresponding energy
expression  takes the form 
\ba
E^{\prime} = -A (m^2 - {N^2 \over 4}) 
= -{A\over 2}(N+1/2) + A (N+1) (v + 1/2)  -A (v+1/2)^2 ~~ ,  
\ea
from which we immediately obtain 
\ba
\omega_e & = &  A (N+1) ~~ ,   \nonumber \\
x_e \omega_e & = &  A ~~ . 
\ea
Thus, in a diatomic molecule the parameters $A$ and $N$ can be 
determined by the spectroscopic constants $\omega_e$ and
$x_e\omega_e$. 

We now  consider the $U_i(2) \supset SU_i(2) \supset SO_i(2)$ 
algebra, which is generated by the set   
$\{ \hat G_i \} \equiv $  $ \{ \hat N_i, \, \hat J_{+,i}, 
\, \hat J_{-,i}, \, \hat J_{0,i} \}$, satisfying the commutation 
relations 
\ba
\, [ \hat J_{0,i}, \hat J_{\pm,i}] \;=\; \pm \hat J_{\pm,i} ~,
\hspace{1cm} 
\, [ \hat J_{+,i}, \hat J_{-,i}] \;=\; 2 \hat J_{0,i} ~,
\hspace{1cm} 
\, [ \hat N_i, \hat J_{\mu,i}] \;=\; 0 ~, \label{jmui}
\ea
with $\mu=\pm,0$.  As mentioned before, 
 for the symmetric irreducible representation
$[N_i,0]$  
of $U_i(2)$ one can show that the Casimir operator is given by
$\vec{J}_i^{\, 2} = \hat N_i(\hat N_i+2)/4$ \cite{diecinueve}, 
from which follows the identification $j_i=N_i/2$. The $SO_i(2)$
label is denoted by $m_i$. 

\section{The Be$_4$ cluster} 

As a specific example, we  
consider the Be$_4$ cluster, which has a tetrahedral shape.  
${\cal D}_{3h}$ molecules can be similarly treated. In the Be$_4$  
case there are six $U_i(2)$ algebras involved ($i=1,\ldots,6$).
In the present approach each relevant interatomic interaction
is associated with a $U_i(2)$ algebra. 
The operators in the model are expressed in terms of the 
generators of these algebras, and the symmetry requirements of the 
tetrahedral group ${\cal T}_d$  can be readily imposed \cite{trece,veinte}. 
The local operators $\{ \hat G_i \}$ 
acting on bond $i$ can be projected to any of the fundamental 
irreps $\Gamma=A_1$, $E$ and  $F_2$.
Using the $\hat J_{\mu,i}$ generators 
we obtain the ${\cal T}_d$ tensors
\ba
\hat T^{\Gamma}_{\mu,\gamma} &=& 
\sum_{i=1}^{6} \, \alpha^{\Gamma}_{\gamma,i} \, \hat J_{\mu,i} ~,
\ea
where $\mu=\pm,0$ and $\gamma$ denotes the component of $\Gamma$. 
The explicit expressions  are given by
\ba
\hat T^{A_1}_{\mu,1} &=& \frac{1}{\sqrt{6}}  
\sum_{i=1}^{6} \, \hat J_{\mu,i} ~, 
\nonumber\\ 
\hat T^{E}_{\mu,1} &=& \frac{1}{2\sqrt{3}} \left( \hat J_{\mu,1} 
+ \hat J_{\mu,2} - 2 \hat J_{\mu,3} + \hat J_{\mu,4} 
- 2 \hat J_{\mu,5} + \hat J_{\mu,6} \right) ~, 
\nonumber\\ 
\hat T^{E}_{\mu,2} &=& \frac{1}{2} \left( \hat J_{\mu,1} 
- \hat J_{\mu,2} - \hat J_{\mu,4} + \hat J_{\mu,6} \right) ~, 
\nonumber\\
\hat T^{F_2}_{\mu,1} &=& \frac{1}{\sqrt{2}} 
\left( \hat J_{\mu,1} - \hat J_{\mu,6} \right) ~,
\nonumber\\
\hat T^{F_2}_{\mu,2} &=& \frac{1}{\sqrt{2}} 
\left( \hat J_{\mu,2} - \hat J_{\mu,4} \right) ~,
\nonumber\\
\hat T^{F_2}_{\mu,3} &=& \frac{1}{\sqrt{2}} 
\left( \hat J_{\mu,3} - \hat J_{\mu,5} \right) ~. \label{tdgen}
\ea
The Hamiltonian operator can be constructed by repeated couplings 
of these tensors to a total symmetry $A_1$, since it must commute 
with all operations in ${\cal T}_d$ \cite{trece}. 

All calculations are carried out in a symmetry-adapted basis, which 
is projected from the local basis
\ba
\begin{array}{ccccccccccccc}
U_1(2) &\otimes& \cdots &\otimes& U_6(2) &\supset& 
SO_1(2) &\otimes& \cdots &\otimes& SO_6(2) &\supset& SO(2) \\
\downarrow && && \downarrow && \downarrow && && 
\downarrow && \downarrow \\ 
| \;\; [N_1] &,& \ldots &,& [N_6] &;& 
v_1 &,& \ldots &,& v_6 &;& \; V \;\; \rangle 
\end{array}
\ea
in which each anharmonic oscillator is well defined. By symmetry  
considerations, $N_i=N$ for the six oscillators, $v_i= N_i/2 - m_i$
denotes the number of quanta in bond $i$ and $V=\sum_i v_i$ is the total 
number of quanta. The local basis states for each 
oscillator are usually written as $|N_i,v_i \rangle$, where 
$v_i=(N_i-2m_i)/2=0,1, \ldots [N_i/2]$ denotes the number of oscillator 
quanta in the $i$-th oscillator.  
The states with one quantum $V=1$ are denoted by 
$| \, i \, \rangle$ with $v_i=1$ and $v_{j \neq i}=0$. Using the same
projection technique as for the generators (\ref{tdgen}), we 
find the six fundamental modes
\ba
| \, ^{1}\phi^{\Gamma}_{\gamma} \rangle  &=& 
\sum_{i=1}^{6} \, \alpha^{\Gamma}_{\gamma,i} \, | \, i \, \rangle ~. 
\label{tdbas}
\ea
The expansion coefficients are the same as in Eq.~(\ref{tdgen}).
The states with a higher number of quanta 
$|^{V}\phi^{\Gamma}_{\gamma} \rangle$ can be  
constructed using the Clebsch-Gordan coefficients of 
${\cal T}_d$ \cite{trece,veinte}. Since all operators are expressed in
terms of powers of the $U_i(2)$ generators, their matrix elements can 
be easily  evaluated in closed form. The symmetry-adapted operators 
of Eq.~(\ref{tdgen}) and symmetry-adapted basis states are the building 
blocks of the model.  

We now proceed to expicitly construct the  Be$_4$   Hamiltonian. 
For interactions that are at most quadratic in the 
generators the  procedure yields 
\ba
\hat H_0 &=& \omega_1 \, \hat{\cal H}_{A_1} 
+ \omega_2 \, \hat{\cal H}_{E} 
+ \omega_3 \, \hat{\cal H}_{F_2} 
+ \alpha_2 \, \hat{\cal V}_{E} 
+ \alpha_3 \, \hat{\cal V}_{F_2} ~, \label{H0}
\ea
with
\ba
\hat{\cal H}_{\Gamma} &=& \frac{1}{2N} \sum_{\gamma} \left( 
  \hat T^{\Gamma}_{-,\gamma} \, \hat T^{\Gamma}_{+,\gamma}
+ \hat T^{\Gamma}_{+,\gamma} \, \hat T^{\Gamma}_{-,\gamma} 
\right) ~,
\nonumber\\
\hat{\cal V}_{\Gamma} &=& \frac{1}{N} \sum_\gamma   
\hat T^{\Gamma}_{0,\gamma} \, \hat T^{\Gamma}_{0,\gamma} ~.
\ea
Note that we have not included  
$\hat{\cal V}_{A_1}$ in $\hat H_0$, since the combination 
\ba
\sum_{\Gamma} \left( \hat{\cal H}_{\Gamma} + \hat{\cal V}_{\Gamma} \right) 
&=& \frac{1}{4N} \sum_{i=1}^{6} \hat N_i(\hat N_i +2) ~, 
\ea
is a constant $3(N+2)/2$~. 
The five interaction terms in Eq.~(\ref{H0}) correspond 
to  linear combinations of the Casimir operators of \cite{trece}. 
However, for a good description of the vibrational energies of Be$_4$ 
it is necessary  to include interactions  which are related 
to the vibrational angular momenta associated with the 
degenerate modes $E$ and $F_2$. These kind of terms is absent in the 
former versions of the model \cite{once,trece}. 
We now proceed to show how they can be obtained in the present formalism. 
In configuration space the vibrational angular momentum operator 
for the $E$ mode is given by \cite{Hecht}
\ba
\hat l^{A_2} &=& -i \left( q^E_1 \, \frac{\partial}{\partial q^E_2}
- q^E_2 \, \frac{\partial}{\partial q^E_1} \right) ~,
\ea
where $q^E_1$ and $q^E_2$ are the  normal coordinates
associated to the  $E$ mode.   
This relation can be transformed to the algebraic space by means of 
the harmonic oscillator operators
\ba
b^{\Gamma \, \dagger}_{\gamma} \;=\; \frac{1}{\sqrt{2}} \left( 
q^{\Gamma}_{\gamma} - \frac{\partial}{\partial q^{\Gamma}_{\gamma}}
\right) ~, \hspace{1cm}
b^{\Gamma}_{\gamma} \;=\; \frac{1}{\sqrt{2}} \left( 
q^{\Gamma}_{\gamma} + \frac{\partial}{\partial q^{\Gamma}_{\gamma}}
\right) ~,
\ea
to obtain
\ba
\hat l^{A_2} &=& -i \left( b^{E \, \dagger}_1 b^{E}_2 - 
b^{E \, \dagger}_2 b^{E}_1 \right) ~.
\label{vibang}
\ea
Here $b^{E}_{\gamma} = \sum_{i} \alpha^E_{\gamma,i} \, b_i$, with a
similar form for $b^{\Gamma \, \dagger}_{\gamma}$, while the
$\alpha^E_{\gamma \, i}$ can be read from Eqs.~(12,13).   
In order to find the algebraic expression for $\hat l^{A_2}$  
 we first  introduce a scale transformation  
\ba
\bar b^{\dagger}_i \;\equiv\; \hat J_{-,i}/\sqrt{N_i} ~,
\hspace{1cm} 
\bar b_i \;\equiv\; \hat J_{+,i}/\sqrt{N_i} ~. \label{subst}
\ea
The relevant commutator can then be expressed as
\ba
[\bar b_i, \bar b_i^\dagger] \;=\; 
\frac{1}{N_i} [ \hat J_{+,i},\hat J_{-,i}] 
\;=\; \frac{1}{N_i} 2\hat J_{0,i} \;=\; 1 - \frac{2 \hat v_i}{N_i} ~,
\ea
where 
\ba
\hat v_i=\frac{\hat N_i}{2}-\hat J_{0,i} ~.
\ea
The other two commutation relations of Eq.~(11) are not modified by 
the scale transformation of Eq.~(22).  
In the harmonic limit, which is defined by $N_i \rightarrow \infty$,
Eq.~(23) reduces to the standard boson commutator  
$[\bar b_i, \bar b^\dagger_i]=1$.  
This limit corresponds to a contraction of $SU(2)$ to the Weyl algebra
and can be used to obtain a geometric interpretation of algebraic   
operators in terms of those in configuration space. 
In the opposite sense, Eq.~(\ref{subst}) provides a procedure to 
construct the  anharmonic representation of harmonic operators 
through the correspondence 
$b^{\dagger}_i \rightarrow \bar b^\dagger_i = 
 \hat J_{-,i}/\sqrt{N_i}$ and 
$b_i \rightarrow \bar b _i = \hat J_{+,i}/\sqrt{N_i}$.
Applying this method to the vibrational angular momentum we find 
\ba 
\hat l^{A_2} &=& -\frac{i}{N} \left( \hat T^E_{-,1} \hat T^E_{+,2} -
\hat T^E_{-,2} \hat T^E_{+,1} \right) ~.
\ea
For the vibrational angular momentum $\hat l^{F_1}_{\gamma}$ 
associated with the $F_2$ mode we find a similar 
expression. The corresponding interactions are 
\ba
\hat H_1 &=& g_{22} \, \hat l^{A_2} \, \hat l^{A_2} 
+ g_{33} \, \sum_{\gamma} \hat l^{F_1}_{\gamma} \, \hat
l^{F_1}_{\gamma} ~. 
\ea
With this method we obtain  
an algebraic realization of arbitrary configuration space
interactions. As a simple example, a one-dimensional harmonic
oscillator Hamiltonian $\hat H_i = (b^\dagger_i b_i + b_i
b^\dagger_i)/2$, transforms into 
\ba
\frac{1}{2N} (\hat J_{-,i} \hat J_{+,i} + \hat J_{+, i} \hat J_{-,i})
= \frac{1}{N} (\hat J^2_i - \hat J^2_{0,i}) = \hat v_i + 1/2 - \frac{
\hat v^2_i}{N} ~,
\ea
where in the last step we used Eq.~(24).  The spectrum of Eq.~(27)
has an anharmonic correction, analogous to the quadratic term in the
Morse potential spectrum.  We are thus substituting harmonic
oscillators by Morse oscillators. 

A more interesting application is to use  
our model to fit the spectroscopic data of several polyatomic
molecules.  In the case of Be$_4$ the energy spectrum  
was analyzed by {\it ab initio} methods in \cite{quince}, where force-field
constants corresponding to an expansion of the potential up to fourth
order in the normal coordinates and momenta were evaluated. We have
generated the {\it ab initio} spectrum up to three quanta using the
analysis in \cite{Hecht}. For the algebraic Hamiltonian we take 
\cite{diecisiete} 
\ba
\hat H &=& 
\omega_1 \, \hat{\cal H}_{A_1} + \omega_2 \, \hat{\cal H}_{E}
+ \omega_3 \, \hat{\cal H}_{F_2} 
+ X_{12} \left( \hat{\cal H}_{A_1} \hat{\cal H}_{E  } \right)
+ X_{13} \left( \hat{\cal H}_{A_1} \hat{\cal H}_{F_2} \right)
+ X_{33} \left( \hat{\cal H}_{F_2} \right)^2
\nonumber\\
&& + g_{33} \, \sum_{\gamma} 
\hat l^{F_1}_{\gamma} \, \hat l^{F_1}_{\gamma} 
+ t_{33} \, \hat{\cal O}_{33} 
+ t_{23} \, \hat{\cal O}_{23} ~,
\ea
The terms $\hat{\cal O}_{33}$ and $\hat{\cal O}_{23}$ represent 
the algebraic form of the corresponding interactions 
in \cite{Hecht} which are responsible for the splitting of the 
vibrational levels in the $(\nu_1,\nu_2^m,\nu_3^l)=(0,0^0,2^2)$ and 
the $(0,1^1,1^1)$ overtones \cite{diecisiete}.  

In Table I we show the the results of a least-square
fit to the vibrational energies of Be$_4$ with 
the Hamiltonian of Eq.~(28).  
The r.m.s. deviation obtained is 2.6 cm$^{-1}$, which can be considered
of spectroscopic quality.  We point out that in
\cite{Hecht,veinticinco}  several
higher order interactions are present which we have  neglected.
Since our model  can be put into a one to one correspondence with the
configuration space calculations, it is in fact possible to improve
the accuracy of the fit considerably, but we have used a simpler
Hamiltonian than the one of \cite{Hecht,veinticinco}.  When no
{\it ab initio} 
calculations are available (or feasible) the present approach can be
used empirically, achieving increasingly good fits by the inclusion
of higher order interactions \cite{diecisiete}.  

We note that the Be$_4$ Hamiltonian of Eq.~(28) preserves the total
number of quanta $V$.  This is a good approximation for this case
according to the analysis of \cite{Hecht,veinticinco}, but it is
known that Fermi  
resonances can occur for certain molecules when the fundamental mode
frequencies are such that $(V,V^\prime)$ states with $V \neq
V^\prime$ are close in energy.  These interactions can be introduced
in the Hamiltonian but the size of the energy matrices grows very
rapidly, so the best way to deal with this problem is through
perturbation theory.  

\section{${\cal D}_{3h}$ triatomic molecules}

For ${\cal D}_{3h}$ molecules we
follow a similar procedure, namely, we construct the
${\cal D}_{3h}$ symmetry-adapted operators and states analogous 
to Eq.~(13,15) and carry out the building up procedure to construct
the Hamiltonian and states with a higher number of quanta with the
appropriate projection operators and Clebsch-Gordan 
coefficients \cite{X3}.  
 
In Table II we 
present the fits to the spectra of Be$_3$, Na$^+_3$ and H$^+_3$ up
to three quanta.  While remarkably accurate descriptions of the
first two molecules can be achieved using a four-parameter
Hamiltonian, we were forced to include four additional higher order
terms in the H$^+_3$ Hamiltonian in order to properly describe this
molecule.  This is in accordance with the work of Carter and
Meyer \cite{dieciseis}, who were forced to include twice as many
terms in the potential energy surface for H$^+_3$ than for the
Na$^+_3$ molecule.  The H$^+_3$ ion is a very ``soft'' molecule
which, due to the light mass of its atomic constituents carries out
large amplitude oscillations from its equilibrium
positions \cite{dieciseis}. 

\section{The methane molecule}

We now turn our attention to the CH$_4$ molecule, for which we shall
make a detailed description. For methane we have 
four $U(2)$ algebras corresponding to the C-H interactions and six more 
representing the H-H couplings. The assignments and the choice of the 
Cartesian coordinate system are the same as in \cite{trece}.
The molecular dynamical group is then given by the product 
\ba
{\cal G} &=& U_1(2) \otimes U_2(2) \otimes \ldots \otimes U_{10}(2) ~.
\ea
The labeling is such that $i=1,\ldots,4$ correspond to 
the C-H couplings while the other values of $i$ are associated with 
H-H interactions.
  Consequently  there are two different boson 
numbers, $N_s$ for the C-H couplings and $N_b$ for the H-H couplings,
which correspond to the stretching and bending modes, respectively.
  The tetrahedral symmetry of methane is taken into account by projecting 
the local operators $\{ \hat G_i \}$, which act on bond $i$, on the 
irreducible representations $\Gamma$ of the tetrahedral group 
${\cal T}_d$. The explicit expressions for the ${\cal T}_d$ tensors 
for the stretching modes are 
\ba
\hat{T}^{A_{1,s}}_{\mu,1} &=& \frac{1}{2}  
\sum_{i=1}^{4} \, \hat J_{\mu,i} ~, 
\nonumber\\ 
\hat{T}^{F_{2,s}}_{\mu,1} &=& \frac{1}{2} 
\left( \hat J_{\mu,1} - \hat J_{\mu,2} 
+ \hat J_{\mu,3} - \hat J_{\mu,4} \right) ~,
\nonumber\\ 
\hat{T}^{F_{2,s}}_{\mu,2} &=& \frac{1}{2} 
\left( \hat J_{\mu,1} - \hat J_{\mu,2} 
- \hat J_{\mu,3} + \hat J_{\mu,4} \right) ~,
\nonumber\\ 
\hat{T}^{F_{2,s}}_{\mu,3} &=& \frac{1}{2} 
\left( \hat J_{\mu,1} + \hat J_{\mu,2} 
- \hat J_{\mu,3} - \hat J_{\mu,4} \right) ~, \label{stretch}
\ea
while for the bending modes we have
\ba
\hat{T}^{A_{1,b}}_{\mu,1} &=& \frac{1}{\sqrt{6}}  
\sum_{i=5}^{10} \, \hat J_{\mu,i} ~, 
\nonumber\\ 
\hat{T}^{E_b}_{\mu,1} &=& \frac{1}{2\sqrt{3}} 
\left( \hat J_{\mu,5} + \hat J_{\mu,6} - 2 \hat J_{\mu,7} 
+ \hat J_{\mu,8} - 2 \hat J_{\mu,9} + \hat J_{\mu,10} \right) ~, 
\nonumber\\ 
\hat{T}^{E_b}_{\mu,2} &=& \frac{1}{2} 
\left( \hat J_{\mu,5} - \hat J_{\mu,6} 
- \hat J_{\mu,8} + \hat J_{\mu,10} \right) ~, 
\nonumber\\
\hat{T}^{F_{2,b}}_{\mu,1} &=& \frac{1}{\sqrt{2}} 
\left( \hat J_{\mu,5} - \hat J_{\mu,10} \right) ~,
\nonumber\\
\hat{T}^{F_{2,b}}_{\mu,2} &=& \frac{1}{\sqrt{2}} 
\left( \hat J_{\mu,6} - \hat J_{\mu,8} \right) ~,
\nonumber\\
\hat{T}^{F_{2,b}}_{\mu,3} &=& \frac{1}{\sqrt{2}} 
\left( \hat J_{\mu,7} - \hat J_{\mu,9} \right) ~. \label{bend}
\ea
As before, the algebraic Hamiltonian can be constructed by 
repeated couplings of these tensors to a total symmetry $A_1$. 

The methane molecule has nine vibrational degrees of freedom. Four of 
them correspond to the fundamental stretching modes ($A_1 \oplus F_2$) 
and the other five to the fundamental bending modes ($E \oplus F_2$) 
\cite{veintiseis}. The projected tensors of Eqs.~(\ref{stretch}) 
and~(\ref{bend}) correspond to ten degrees of freedom, four of which 
($A_1 \oplus F_2$) are related to stretching modes and six 
($A_1 \oplus E \oplus F_2$) to the bendings. Consequently we can 
identify the tensor $\hat{T}^{A_{1,b}}_{\mu,1}$ as the operator associated 
to a spurious mode. This identification makes it possible to eliminate the 
spurious states {\em exactly}. This is achieved by (i) ignoring the 
$\hat{T}^{A_{1,b}}_{\mu,1}$ tensor in the construction of the Hamiltonian, 
and (ii) diagonalizing this Hamiltonian in a symmetry-adapted basis from 
which the spurious mode has been removed following the procedure of 
\cite{trece}. We note that the condition on the Hamiltonian that was used in 
\cite{trece} to exclude the spurious contributions, does not automatically 
hold for states with higher number of quanta.

According to the above procedure, we now construct the ${\cal T}_d$ 
invariant interactions that are at most quadratic in the generators 
and conserve the total number of quanta 
\ba
\hat{\cal H}_{\Gamma_x} &=& \frac{1}{2N_{x}} \sum_{\gamma} \left( 
  \hat T^{\Gamma_x}_{-,\gamma} \, \hat T^{\Gamma_x}_{+,\gamma}
+ \hat T^{\Gamma_x}_{+,\gamma} \, \hat T^{\Gamma_x}_{-,\gamma} 
\right) ~,
\nonumber\\
\hat{\cal V}_{\Gamma_x} &=& \frac{1}{N_{x}} \sum_{\gamma} \,
\hat T^{\Gamma_x}_{0,\gamma} \, \hat T^{\Gamma_x}_{0,\gamma} ~.
\label{hv}
\ea
Here $\Gamma=A_1$, $F_2$ for the stretching vibrations $x=s$ and 
$\Gamma=E$, $F_2$ for the bending vibrations $x=b$. In addition 
there are two stretching-bending interactions
\ba
\hat{\cal H}_{sb} &=& \frac{1}{2\sqrt{N_sN_b}} \sum_{\gamma} \left( 
  \hat T^{F_{2,s}}_{-,\gamma} \, \hat T^{F_{2,b}}_{+,\gamma}
+ \hat T^{F_{2,s}}_{+,\gamma} \, \hat T^{F_{2,b}}_{-,\gamma} \right) ~,
\nonumber\\
\hat{\cal V}_{sb} &=& \frac{1}{\sqrt{N_sN_b}} \sum_{\gamma} \, 
\hat T^{F_{2,s}}_{0,\gamma} \, \hat T^{F_{2,b}}_{0,\gamma} ~.
\label{hvsb}
\ea
The zeroth order vibrational Hamiltonian is now written as
\ba
\hat H_0 &=& \omega_1 \, \hat{\cal H}_{A_{1,s}} 
   + \omega_2 \, \hat{\cal H}_{E_b} 
   + \omega_3 \, \hat{\cal H}_{F_{2,s}} 
   + \omega_4 \, \hat{\cal H}_{F_{2,b}} 
   + \omega_{34} \, \hat{\cal H}_{sb} 
\nonumber\\
&& + \alpha_2 \, \hat{\cal V}_{E_b}  
   + \alpha_3 \, \hat{\cal V}_{F_{2,s}} 
   + \alpha_4 \, \hat{\cal V}_{F_{2,b}} 
   + \alpha_{34} \, \hat{\cal V}_{sb} ~. \label{h0}
\ea
The interaction $\hat{\cal V}_{A_{1,s}}$ has not been included since,
in analogy to Eq.~(18), the combination
\ba
\sum_{\Gamma} \left( \hat{\cal H}_{\Gamma_s} 
+ \hat{\cal V}_{\Gamma_s} \right) &=& \frac{1}{4N_s}
\sum_{i=1}^{4} \hat N_i(\hat N_i+2) ~,
\ea
corresponds to a constant $N_s+2$. A similar relation holds for the 
bending interactions, but in this case the interaction 
$\hat{\cal V}_{A_{1,b}}$ has already been excluded in order to remove 
the spurious $A_1$ bending mode. The subscripts of the parameters 
correspond to the $(\nu_1,\nu_2^{l_2},\nu_3^{l_3},\nu_4^{l_4})$ 
labeling of a set of basis states for 
the vibrational levels of CH$_4$. Here $\nu_1$, $\nu_2$, $\nu_3$ and 
$\nu_4$ denote the number of quanta in the $A_{1,s}$, $E_b$, 
$F_{2,s}$ and $F_{2,b}$ modes, respectively. The labels $l_i$ are 
related to the vibrational angular momentum associated with degenerate 
vibrations. The allowed values are $l_i=\nu_i,\nu_i-2,\ldots,1$ or 0 
for $\nu_i$ odd or even \cite{veintiseis}.

In the harmonic limit the interactions of Eqs.~(32) 
and~(33) again attain a particularly simple form, which can be 
directly related to configuration space interactions. 
This limit is obtained, as before, by rescaling $\hat J_{+,i}$ and
$\hat J_{-,i}$  
by $\sqrt{N_i}$ and taking $N_i \rightarrow \infty$, so that
\ba
\lim_{N_i \rightarrow \infty} \frac{\hat J_{+,i}}{\sqrt{N_i}}
&=& b_i ~,
\nonumber\\
\lim_{N_i \rightarrow \infty} \frac{\hat J_{-,i}}{\sqrt{N_i}}
&=& b_i^{\dagger} ~,
\nonumber\\
\lim_{N_i \rightarrow \infty} \frac{1}{N_i} [ \hat J_{+,i},\hat J_{-,i}] 
&=& \lim_{N_i \rightarrow \infty} \frac{2\hat J_{0,i}}{N_i} 
\;=\; 1 ~.
\ea
where the operators $b_i$ and $b_j^{\dagger}$ satisfy the standard boson 
commutation relation $[b_i,b^\dagger_j]=\delta_{ij}$. Applying the 
harmonic limit to the interactions of Eqs.~(32) and~(33) 
we obtain
\ba
\lim_{N_{x} \rightarrow \infty} \, \hat{\cal H}_{\Gamma_x} 
&=& \frac{1}{2} \sum_{\gamma} \left( 
  b^{\Gamma_x \, \dagger}_{\gamma} \, b^{\Gamma_x}_{\gamma}
+ b^{\Gamma_x}_{\gamma} \, b^{\Gamma_x \, \dagger}_{\gamma} \right) ~,
\nonumber\\
\lim_{N_{x} \rightarrow \infty} \, \hat{\cal V}_{\Gamma_x} &=& 0 ~,
\nonumber\\
\lim_{N_s, N_b \rightarrow \infty} \, \hat{\cal H}_{sb} 
&=& \frac{1}{2} \sum_{\gamma} \left( 
  b^{F_{2,s} \, \dagger}_{\gamma} \, b^{F_{2,b}}_{\gamma}
+ b^{F_{2,s}}_{\gamma} \, b^{F_{2,b} \, \dagger}_{\gamma} \right) ~,
\nonumber\\
\lim_{N_s, N_b \rightarrow \infty} \, \hat{\cal V}_{sb} &=& 0 ~.
\label{harlim}
\ea
Here the operators $b^{\Gamma_x \, \dagger}_{\gamma}$ are given in 
terms of the local boson operators $b^{\dagger}_i$ through 
the coefficients $\alpha^{\Gamma_x}_{\gamma,i}$ given in 
Eqs.~(30,31)
\ba
b^{\Gamma_x \, \dagger}_{\gamma} &=& 
\sum_{i=1}^{10} \, \alpha^{\Gamma_x}_{\gamma,i} \, b^{\dagger}_i ~,
\ea
with a similar relation for the annihilation operators. From 
Eq.~(37) the physical interpretation of the interactions 
is immediate. The $\hat{\cal H}_{\Gamma_x}$ terms represent the 
anharmonic counterpart of the harmonic interactions, while 
the $\hat{\cal V}_{\Gamma_x}$ terms are purely anharmonic contributions 
which vanish in the harmonic limit. In an application to the ozone 
molecule it was found that these terms can account for the strong 
anharmonicities and incorporate the effect of Darling-Dennison 
type couplings \cite{ozono}.

The zeroth order Hamiltonian of Eq.~(34) is not sufficient to 
obtain a high-quality fit of the vibrations of methane. Several 
physically meaningful interaction terms that are essential for such 
a fit are not present 
in Eq.~(34). They arise in our model as higher order interactions. 
It is an advantage of the symmetry-adapted model 
that the various interaction terms have a direct physical interpretation 
and a specific action on the various modes \cite{diecisiete}. 
Hence the addition of 
higher order terms and anharmonicities can be done in a systematic way.
For the study of the 
vibrational excitations of methane we use the ${\cal T}_d$ invariant 
Hamiltonian \cite{metano} 
\ba
\hat H &=& \omega_1 \, \hat{\cal H}_{A_{1,s}} 
         + \omega_2 \, \hat{\cal H}_{E_b} 
         + \omega_3 \, \hat{\cal H}_{F_{2,s}} 
         + \omega_4 \, \hat{\cal H}_{F_{2,b}} 
         + \alpha_3 \, \hat{\cal V}_{F_{2,s}} 
\nonumber\\
&& + X_{11} \left( \hat{\cal H}_{A_{1,s}} \right)^2
   + X_{22} \left( \hat{\cal H}_{E_{b}  } \right)^2
   + X_{33} \left( \hat{\cal H}_{F_{2,s}} \right)^2
   + X_{44} \left( \hat{\cal H}_{F_{2,b}} \right)^2
\nonumber\\
&& + X_{12} \left( \hat{\cal H}_{A_{1,s}} \, \hat{\cal H}_{E_b    } \right)
   +  X_{14} \left( \hat{\cal H}_{A_{1,s}} \, \hat{\cal H}_{F_{2,b}} \right)
\nonumber\\
&& + X_{23} \left( \hat{\cal H}_{E_b    } \, \hat{\cal H}_{F_{2,s}} \right) 
   + X_{24} \left( \hat{\cal H}_{E_b    } \, \hat{\cal H}_{F_{2,b}} \right) 
   + X_{34} \left( \hat{\cal H}_{F_{2,s}} \, \hat{\cal H}_{F_{2,b}} \right) 
\nonumber\\
&& + g_{22} \, \left( \hat l^{A_2} \right)^2  
   + g_{33} \, \sum_{\gamma} \hat l^{F_1}_{s,\gamma} \, 
                             \hat l^{F_1}_{s,\gamma} 
   + g_{44} \, \sum_{\gamma} \hat l^{F_1}_{b,\gamma} \, 
                             \hat l^{F_1}_{b,\gamma} 
   + g_{34} \, \sum_{\gamma} \hat l^{F_1}_{s,\gamma} \, 
                             \hat l^{F_1}_{b,\gamma} 
\nonumber\\
&& + t_{33} \, \hat{\cal O}_{ss}
   + t_{44} \, \hat{\cal O}_{bb}
   + t_{34} \, \hat{\cal O}_{sb}
   + t_{23} \, \hat{\cal O}_{2s}
   + t_{24} \, \hat{\cal O}_{2b} ~. \label{hamilt}
\ea
The interpretation of the $\omega_i$ and $\alpha_3$ terms follows from 
Eq.~(37). The $X_{ij}$ terms are quadratic in the operators 
$\hat{\cal H}_{\Gamma_x}$ and hence represent anharmonic vibrational 
interactions. The $g_{ij}$ terms are related to the 
vibrational angular momenta associated with the degenerate
vibrations.  As  mentioned before, 
these interactions, which are fundamental to describe molecular systems with 
a high degree of symmetry, are absent in previous versions of the vibron 
model in which the interaction terms are expressed in terms of Casimir 
operators and products thereof \cite{once,trece}. They give 
rise to a splitting of vibrational levels with the same values of 
$(\nu_1,\nu_2,\nu_3,\nu_4)$ but with different $l_2$, $l_3$ and/or $l_4$. 
Their algebraic realization is given by
\ba 
\hat l^{A_2} &=& -i \, \sqrt{2} \frac{1}{N_b} 
[ \hat T^{E_b}_{-} \times \hat T^{E_b}_{+} ]^{A_2} ~,
\nonumber\\
\hat l^{F_1}_{x,\gamma} &=& +i \, \sqrt{2} \frac{1}{N_x} 
[ \hat T^{F_{2,x}}_{-} \times \hat T^{F_{2,x}}_{+} ]^{F_1}_{\gamma} ~.
\label{gij}
\ea
The square brackets in Eq.~(40) denote the tensor coupling 
under the point group ${\cal T}_d$
\ba
[ \hat T^{\Gamma_1} \times \hat T^{\Gamma_2} ]^{\Gamma}_{\gamma} 
&=& \sum_{\gamma_1,\gamma_2} 
C(\Gamma_1,\Gamma_2,\Gamma ; \gamma_1,\gamma_2,\gamma) \,
\hat T^{\Gamma_1}_{\gamma_1} \, \hat T^{\Gamma_2}_{\gamma_2} ~,
\label{tensor}
\ea
where the expansion coefficients are the Clebsch-Gordan coefficients for 
${\cal T}_d$ \cite{trece,veinte}. In the harmonic limit the expectation 
value of the diagonal terms in Eq.~(39) leads to the familiar 
Dunham expansion \cite{veintiseis} 
\ba
\sum_i \omega_i \, (v_i + \frac{d_i}{2}) + \sum_{j \geq i} \sum_i 
X_{ij} \, (v_i + \frac{d_i}{2}) (v_j + \frac{d_j}{2})
+ \sum_{j \geq i} \sum_i g_{ij} \, l_i l_j ~. \label{Dunham}
\ea
Here $d_i$ is the degeneracy of the vibration. 
The $t_{ij}$ terms in Eq.~(39) 
give rise to further splittings of the vibrational levels 
$(\nu_1,\nu_2,\nu_3,\nu_4)$ into its possible sublevels. 
They can be expressed in terms of the tensor operators of 
Eqs.~(30) and~(31) as
\ba
\hat{\cal O}_{xy} &=& \frac{1}{N_x N_y}
\left( 6 \sum_{\gamma} 
[ \hat T^{F_{2,x}}_- \times \hat T^{F_{2,y}}_- ]^{E}_{\gamma} \,
[ \hat T^{F_{2,y}}_+ \times \hat T^{F_{2,x}}_+ ]^{E}_{\gamma} 
-4 \sum_{\gamma}
[ \hat T^{F_{2,x}}_- \times \hat T^{F_{2,y}}_- ]^{F_2}_{\gamma} \,
[ \hat T^{F_{2,y}}_+ \times \hat T^{F_{2,x}}_+ ]^{F_2}_{\gamma} \right) ~, 
\nonumber\\
\hat{\cal O}_{2x} &=& \frac{1}{N_b N_x} 
\left( 8 \sum_{\gamma} 
[ \hat T^{E_b}_- \times \hat T^{F_{2,x}}_- ]^{F_1}_{\gamma} \,
[ \hat T^{E_b}_+ \times \hat T^{F_{2,x}}_+ ]^{F_1}_{\gamma} 
-8 \sum_{\gamma}
[ \hat T^{E_b}_- \times \hat T^{F_{2,x}}_- ]^{F_2}_{\gamma} \,
[ \hat T^{E_b}_+ \times \hat T^{F_{2,x}}_+ ]^{F_2}_{\gamma} \right) ~.
\label{tij}
\ea
In the harmonic limit the $t_{ij}$ terms have the same interpretation 
as in \cite{Hecht}. The $\hat{\cal O}_{ss}$, $\hat{\cal O}_{bb}$ and 
$\hat{\cal O}_{sb}$ terms give rise to a splitting of the $E$ and $F_2$ 
vibrations belonging to the 
$(\nu_1,\nu_2^{l_2},\nu_3^{l_3},\nu_4^{l_4})=(0,0^0,2^2,0^0)$, 
$(0,0^0,0^0,2^2)$ and $(0,0^0,1^1,1^1)$ levels, respectively. 
Similarly, the $\hat{\cal O}_{2s}$ and $\hat{\cal O}_{2b}$ 
terms split the $F_1$ and $F_2$ vibrations belonging to the 
$(0,1^1,1^1,0^0)$ and $(0,1^1,0^0,1^1)$ overtones, respectively. 

The Hamiltonian of Eq.~(\ref{hamilt}) involves 23 interaction 
strengths and the two boson numbers, $N_s$ and $N_b$. The vibron 
number associated with the stretching vibrations is determined 
from the spectroscopic constants $\omega_e$ and $x_e \omega_e$ 
for the CH molecule to be $N_s=43$ \cite{diecinueve}. The vibron number 
for the bending vibrations, which are far more harmonic than the 
stretching vibrations, is taken to be $N_b=150$. We have carried out 
a least-square fit to the vibrational spectrum of methane including 
44 experimental energies from \cite{ch4exp}. 
The values of the fitted parameters 
are presented in the second column of Table~\ref{fit} (Fit 1). 
In Table~\ref{CH4} we compare the results of our calculation with the 
experimentally observed energies. All predicted levels up to $V=3$ 
quanta are included. We find an overall fit to the observed levels 
with a r.m.s. deviation which is an order of magnitude better than 
in previous studies. Whereas the r.m.s. deviations of 
standard vibron model studies in \cite{trece} and 
\cite{veintiocho} are 12.16 and 11.61 cm$^{-1}$ for 19 energies, 
we find a value of 1.16 cm$^{-1}$ for 44 energies. 

In order to address the importance of the $\alpha_3$ term, which is 
completely anharmonic in origin and vanishes in the harmonic limit 
(see Eq.~(\ref{harlim})), we have carried out a calculation without 
this term (Fit 2 of Table~\ref{fit}). With one less interaction term 
the r.m.s. deviation increases from 1.16 to 4.49 cm$^{-1}$. This shows 
the importance of the $\alpha_3$ term to obtain an accurate description 
of the anharmonicities that are present in the data. Similarly, the 
third calculation (Fit 3) shows that, in a fit without the $t_{ij}$ 
terms, the r.m.s. deviation increases from 1.16 to 7.81 cm$^{-1}$. 

In addition, we have carried out a fit in the harmonic limit 
($N_s$, $N_b \rightarrow \infty$). In this limit the $\alpha_3$ term 
vanishes and the algebraic Hamiltonian of Eq.~(\ref{hamilt}) 
reduces to the vibrational Hamiltonian of \cite{Hecht}, 
the harmonic frequencies $\omega_i$ and anharmonic constants 
$X_{ij}$, $g_{ij}$ and $t_{ij}$ having the same meaning. 
The r.m.s. deviation increases to 20.42 cm$^{-1}$ (Fit 4). When we 
use the vibrational Hamiltonian of \cite{Hecht}, which contains 
one additional interaction (called $s_{34}$ in Table~\ref{fit}) 
the quality of the fit does not improve. 
In fact, the r.m.s. deviation increases to 20.90 cm$^{-1}$ (Fit 5). 
The importance of the nondiagonal elements is demonstrated in a 
calculation (Fit 6), in which the data are fitted 
by the (diagonal) Dunham expansion of Eq.~(\ref{Dunham}). For this 
case the r.m.s. deviation is 21.00 cm$^{-1}$, almost the same 
value as for the other two calculations in the harmonic limit. 
The small differences in the parameter values and the r.m.s. deviation 
of Fits 4 and 6 show that, unlike for finite $N_s$ and $N_b$ 
(Fits 1 and 3), in the harmonic limit the nondiagonal 
contributions from the $t_{ij}$ terms are not very important. 

A comparison between the parameter values and the r.m.s. deviations 
of Fits 1--6 in Table~\ref{fit} shows that the 
$\alpha_3$ term and the anharmonic effects in the interaction 
terms of Eq.~(\ref{hamilt}) can only be compensated for in part by the 
anharmonicity constants $X_{ij}$. The r.m.s. deviation increases 
from 1.16 to 4.49 and 20.42 cm$^{-1}$ for Fits 1, 2 and 4, respectively.

\section{Summary and conclusions}

In summary, in this paper we have studied the vibrational excitations 
of several molecules in a symmetry-adapted algebraic model.  In
particular, for the  methane molecule we  find an overall fit 
to the 44 observed levels with a r.m.s. deviation of 1.16 cm$^{-1}$, 
which can be considered of spectroscopic quality. 
We pointed out that for this calculation the $\alpha_3$
term in Eq.~(39) in combination with  
the anharmonic effects in the other interaction terms plays a 
crucial role in obtaining a fit of this quality. Purely anharmonic terms 
of this sort arise naturally in the symmetry-adapted algebraic model, 
but vanish in the harmonic limit. Physically, these contributions arise
from the anharmonic character of the interatomic interactions, and seem 
to play an important role when dealing with molecular 
anharmonicities, especially at higher number of quanta.
This conclusion is supported by  our other applications of the 
symmetry-adapted model to the Be$_4$ cluster \cite{diecisiete} and  
the H$_3^+$, Be$_3$ and Na$_3^+$ molecules \cite{X3}, as well as 
our study of two isotopes of the ozone molecule \cite{ozono}. 

These studies suggest that the symmetry-adapted algebraic model 
provides a numerically efficient tool to study molecular 
vibrations with high precision. The main difference with other 
methods is the use of symmetry-adapted tensors in the construction 
of the Hamiltonian. In this approach, the interactions can be constructed 
in a systematic way, each term has a direct physical interpretation, 
and spurious modes can be eliminated exactly. 
It will be important to further explore the 
scope and applicability of the present approach. 
A more extensive study of methane including rotation-vibration 
couplings, states with a higher number of quanta and transition 
intensities is in progress.  

\section*{Acknowledgements}

We thank Prof. J.C. Hilico for his interest and for making available 
to us his compilation of observed level energies. 
This work was supported in part by the 
European Community under contract nr. CI1$^{\ast}$-CT94-0072, 
DGAPA-UNAM under project IN105194, CONACyT-M\'exico under project 
400340-5-3401E and Spanish DGCYT under project PB95-0533.

\clearpage
\begin{table}
\centering
\caption[]{\small 
Fit to ab initio \cite{veinticinco} calculations for Be$_4$. The
values of the parameters are 
$\omega_1=636$, $\omega_2=453$, $\omega_3=532$, 
$X_{33}=44.276$, $X_{12}= 4.546$, $X_{13}=-2.539$, 
$g_{33}=-15.031$, $t_{33}=-1.679$ and $t_{23}=-1.175$. 
The total number of bosons is $N=44$. 
The parameters and energies are given in cm$^{-1}$.
\normalsize}
\vspace{10pt} \label{Be4}
\begin{tabular}{cclcc|cclcc}
\hline
& & & & & & & & & \\
$V$ & $(\nu_1,\nu_2^m,\nu_3^l)$ & $\Gamma$ 
& Ab initio & Fit &
$V$ & $(\nu_1,\nu_2^m,\nu_3^l)$ & $\Gamma$ 
& Ab initio & Fit \\
& & & $N \rightarrow \infty$ & $N=44$ & 
& & & $N \rightarrow \infty$ & $N=44$ \\
& & & & & & & & & \\
\hline
& & & & \\
1 & $(1,0^0,0^0)$     & $A_1$ &  638.6 &  637.0 & 
3 & $(1,0^0,2^0)$     & $A_1$ & 2106.8 & 2105.6 \\
  & $(0,1^1,0^0)$     & $E$   &  453.6 &  455.0 & 
  & $(1,0^0,2^2)$     & $E$   & 2000.1 & 1999.8 \\
  & $(0,0^0,1^1)$     & $F_2$ &  681.9 &  678.2 & 
  &                   & $F_2$ & 2056.8 & 2052.8 \\
2 & $(2,0^0,0^0)$     & $A_1$ & 1271.0 & 1269.2 & 
  & $(0,3^1,0^0)$     & $E$   & 1341.3 & 1343.7 \\
  & $(1,1^1,0^0)$     & $E$   & 1087.1 & 1087.0 & 
  & $(0,3^3,0^0)$     & $A_1$ & 1355.5 & 1352.5 \\
  & $(1,0^0,1^1)$     & $F_2$ & 1312.6 & 1308.3 & 
  &                   & $A_2$ & 1355.5 & 1354.4 \\
  & $(0,2^0,0^0)$     & $A_1$ &  898.3 &  901.4 & 
  & $(0,2^{0,2},1^1)$ & $F_2$ & 1565.5 & 1565.7 \\
  & $(0,2^2,0^0)$     & $E$   &  905.4 &  906.1 & 
  &                   & $F_2$ & 1584.4 & 1583.1 \\
  & $(0,1^1,1^1)$     & $F_1$ & 1126.7 & 1125.1 & 
  & $(0,2^2,1^1)$     & $F_1$ & 1578.5 & 1578.0 \\
  &                   & $F_2$ & 1135.5 & 1134.1 & 
  & $(0,1^1,2^{0,2})$ & $E$   & 1821.4 & 1821.6 \\
  & $(0,0^0,2^0)$     & $A_1$ & 1484.0 & 1483.0 & 
  &                   & $E$   & 1929.5 & 1929.0 \\
  & $(0,0^0,2^2)$     & $E$   & 1377.3 & 1373.9 & 
  & $(0,1^1,2^2)$     & $A_2$ & 1813.3 & 1813.1 \\
  &                   & $F_2$ & 1434.1 & 1429.6 & 
  &                   & $A_1$ & 1830.8 & 1831.7 \\
3 & $(3,0^0,0^0)$     & $A_1$ & 1897.0 & 1896.7 & 
  &                   & $F_2$ & 1874.4 & 1873.2 \\
  & $(2,1^1,0^0)$     & $E$   & 1714.3 & 1714.3 & 
  &                   & $F_1$ & 1883.2 & 1883.0 \\
  & $(2,0^0,1^1)$     & $F_2$ & 1937.0 & 1933.7 & 
  & $(0,0^0,3^{1,3})$ & $F_2$ & 2136.5 & 2134.2 \\
  & $(1,2^0,0^0)$     & $A_1$ & 1526.6 & 1529.2 & 
  &                   & $F_2$ & 2327.3 & 2326.9 \\
  & $(1,2^2,0^0)$     & $E$   & 1533.7 & 1532.8 & 
  & $(0,0^0,3^3)$     & $F_1$ & 2199.8 & 2197.1 \\
  & $(1,1^1,1^1)$     & $F_1$ & 1752.2 & 1749.7 & 
  &                   & $A_1$ & 2256.5 & 2254.4 \\
  &                   & $F_2$ & 1761.0 & 1759.8 & 
  &                   &       &        &        \\
& & & & & & & & & \\
\hline
\end{tabular}
\end{table}

\clearpage
\begin{table}
\centering
\caption[]{\small Least-square energy fit for the vibrational 
excitations of H$^+_3$, Be$_3$ and Na$^+_3$. The energy differences 
$\Delta E = E_{th} - E_{exp}$ are given in cm$^{-1}$.
\normalsize}
\vspace{10pt} 
\begin{tabular}{cccrrr}
\hline
& & & & & \\
& & & H$^+_3$ & Be$_3$ & Na$^+_3$ \\
$V$ & $(\nu_1,\nu_2^l)$ & $\Gamma$ & $\Delta E$ & $\Delta E$ & $\Delta E$ \\ 
& & & & & \\
\hline
& & & & & \\
1 & $(0,1^1)$ & $E$   & -1.55 &  0.51 &  0.93 \\ 
  & $(1,0^0)$ & $A_1$ &  0.42 &  0.02 &  1.95 \\ 
& & & & &  \\
2 & $(0,2^0)$ & $A_1$ &  7.48 & -0.74 &  0.37 \\ 
  & $(0,2^2)$ & $E$   & -5.69 &  0.17 &  0.84 \\
  & $(1,1^1)$ & $E$   & -0.61 &  0.82 &  1.68 \\
  & $(2,0^0)$ & $A_1$ & -0.11 & -0.04 &  1.26 \\
& & & & &  \\ 
3 & $(0,3^1)$ & $E$   & -4.46 & -2.05 & -1.19 \\ 
  & $(0,3^3)$ & $A_1$ &  3.18 & -1.23 & -0.34 \\
  & $(0,3^3)$ & $A_2$ &  2.44 &  0.61 & -0.33 \\
  & $(1,2^0)$ & $A_1$ &  0.66 &  1.90 & -0.01 \\
  & $(1,2^2)$ & $E$   & -5.00 & -1.36 &  0.34 \\
  & $(2,1^1)$ & $E$   &  4.07 &  0.79 & -0.19 \\
  & $(3,0^0)$ & $A_1$ & -1.23 & -1.66 & -2.06 \\
& & & & & \\
\hline 
& & & & & \\ 
& & r.m.s. &  5.84 &  1.35 &  1.33 \\
& & & & & \\
\multicolumn{3}{r} {Parameters} & 8 & 4 & 4 \\
& & & & & \\
\hline
\end{tabular}
\end{table}

\clearpage
\begin{table}
\centering
\caption[]{\small
Parameters in cm$^{-1}$ obtained in the fits to the vibrational 
energies of CH$_4$.
\normalsize}
\vspace{15pt} \label{fit}
\begin{tabular}{c|rrrrrr}
\hline
& & & & & & \\
Parameter & Fit 1 & Fit 2 & Fit 3 & Fit 4 & Fit 5 & Fit 6 \\
& & & & & & \\
\hline
& & & & & & \\
$N_s$      &      43 &      43 &      43 & $\infty$ & $\infty$ & $\infty$ \\
$N_b$      &     150 &     150 &     150 & $\infty$ & $\infty$ & $\infty$ \\
$\omega_1$ & 2977.60 & 2966.17 & 2970.03 & 2967.40  & 2966.81  & 2969.75  \\
$\omega_2$ & 1554.83 & 1549.96 & 1550.88 & 1558.38  & 1558.51  & 1548.30  \\
$\omega_3$ & 3076.45 & 3076.41 & 3079.33 & 3081.34  & 3082.00  & 3060.48  \\
$\omega_4$ & 1332.22 & 1329.52 & 1337.69 & 1337.51  & 1337.54  & 1338.64  \\
$\alpha_3$ &  582.87 &      -- &  480.33 &      --  &      --  &      --  \\
$X_{11}$   &    3.69 &   10.23 &    6.06 & --21.30  & --21.19  & --21.59  \\
$X_{22}$   &    1.30 &    1.32 &    1.37 &  --1.17  &  --1.17  &  --0.28  \\
$X_{33}$   &    5.43 &    6.97 &    5.34 & --10.79  & --11.12  &  --7.89  \\
$X_{44}$   &  --3.47 &  --3.64 &  --4.41 &  --6.26  &  --6.27  &  --7.26  \\
$X_{12}$   &  --3.60 &  --0.94 &  --1.47 &  --3.39  &  --3.28  &  --2.80  \\
$X_{13}$   &      -- &      -- &      -- &      --  &      --  &      --  \\
$X_{14}$   &  --2.86 &  --0.49 &  --2.30 &  --3.10  &  --3.00  &  --4.48  \\
$X_{23}$   & --11.14 &  --8.75 & --10.68 &  --7.97  &  --8.10  &  --3.80  \\
$X_{24}$   &    1.00 &    0.91 &    2.03 &  --5.37  &  --5.37  &  --4.74  \\
$X_{34}$   &  --5.60 &  --3.97 &  --6.50 &  --3.46  &  --3.50  &  --1.21  \\
$g_{22}$   &  --0.46 &  --0.46 &  --0.41 &    0.37  &    0.37  &  --0.62  \\
$g_{33}$   &    0.19 &  --1.23 &    0.25 &  --4.35  &  --4.22  &  --5.49  \\
$g_{44}$   &    4.07 &    4.11 &    3.79 &    4.98  &    4.98  &    5.30  \\
$g_{34}$   &  --0.65 &  --0.72 &  --0.56 &  --0.74  &  --0.87  &  --0.67  \\
$t_{33}$   &    0.40 &    0.16 &      -- &  --1.25  &  --1.27  &      --  \\
$t_{44}$   &    1.00 &    1.00 &      -- &    0.56  &    0.56  &      --  \\
$t_{34}$   &    0.21 &    0.24 &      -- &    0.24  &    0.25  &      --  \\
$t_{23}$   &  --0.39 &  --0.39 &      -- &  --0.39  &  --0.39  &      --  \\
$t_{24}$   &    0.13 &    0.13 &      -- &    0.91  &    0.91  &      --  \\
$s_{34}$   &      -- &      -- &      -- &      --  &  --0.11  &      --  \\
& & & & & & \\
\hline
& & & & & & \\
r.m.s. & 1.16 & 4.49 & 7.81 & 20.42 & 20.90 & 21.00 \\
& & & & & & \\
\hline
\end{tabular}

\end{table}

\clearpage
\begin{table}
\centering
\caption[]{\small
Fit to vibrational excitations of CH$_4$. The values of the parameters 
are given in the second column of Table\ref{fit}. Here 
$\Delta E=E_{cal}-E_{exp}$. The experimental energies are taken from 
\cite{ch4exp}. 
The wave numbers are given in cm$^{-1}$. 
\normalsize}
\footnotesize
\vspace{15pt} \label{CH4}
\begin{tabular}{lcllr|lcllr}
\hline
& & & & & & & & & \\
$\Gamma$ & $(\nu_1,\nu_2,\nu_3,\nu_4)$ & 
$E_{cal}$ & $E_{exp}$ & $\Delta E$ & 
$\Gamma$ & $(\nu_1,\nu_2,\nu_3,\nu_4)$ & 
$E_{cal}$ & $E_{exp}$ & $\Delta E$ \\
& & & & & & & & & \\
\hline
& & & & & & & & & \\
$A_1$ & (1000) & 2916.32 & 2916.48 & --0.16 & 
      & (0111) & 5844.98 &         &        \\
$E$   & (0100) & 1533.46 & 1533.33 &   0.13 &
      & (1200) & 5974.81 &         &        \\
$F_2$ & (0001) & 1309.86 & 1310.76 & --0.90 &
      & (1011) & 7147.49 &         &        \\
      & (0010) & 3018.09 & 3019.49 & --1.40 &
      & (0021) & 7303.38 &         &        \\
      &        &         &         &        &
      & (2100) & 7315.60 &         &        \\
$A_1$ & (0002) & 2587.77 & 2587.04 &   0.73 &
      & (0120) & 7479.48 &         &        \\
      & (0200) & 3063.66 & 3063.65 &   0.01 &
      & (0120) & 7557.17 &         &        \\
      & (0011) & 4323.81 & 4322.72 &   1.09 &
      & (1020) & 8833.05 &         &        \\
      & (2000) & 5790.13 & 5790    &   0.13 &
$F_1$ & (0003) & 3920.46 & 3920.50 & --0.04 \\
      & (0020) & 5966.57 & 5968.1  & --1.53 &
      & (0102) & 4128.38 & 4128.57 & --0.19 \\
$E$   & (0002) & 2624.14 & 2624.62 & --0.48 &
      & (0201) & 4364.39 & 4363.31 &   1.08 \\
      & (0200) & 3065.22 & 3065.14 &   0.08 &
      & (0012) & 5620.08 &         &        \\
      & (0011) & 4323.09 & 4322.15 &   0.94 &
      & (0012) & 5630.76 &         &        \\
      & (1100) & 4446.41 & 4446.41 &   0.00 &
      & (1101) & 5755.58 &         &        \\
      & (0020) & 6045.03 & 6043.8  &   1.23 &
      & (0111) & 5829.79 &         &        \\
$F_1$ & (0101) & 2845.35 & 2846.08 & --0.73 &
      & (0111) & 5848.94 &         &        \\
      & (0011) & 4323.15 & 4322.58 &   0.57 &
      & (0210) & 6061.57 &         &        \\
      & (0110) & 4537.57 & 4537.57 &   0.00 &
      & (1011) & 7147.53 &         &        \\
$F_2$ & (0002) & 2612.93 & 2614.26 & --1.33 &
      & (0021) & 7303.29 &         &        \\
      & (0101) & 2830.61 & 2830.32 &   0.29 &
      & (0021) & 7343.21 &         &        \\
      & (1001) & 4223.46 & 4223.46 &   0.00 &
      & (1110) & 7361.79 &         &        \\
      & (0011) & 4321.02 & 4319.21 &   1.81 &
      & (0120) & 7518.70 &         &        \\
      & (0110) & 4543.76 & 4543.76 &   0.00 &
      & (0030) & 8947.65 & 8947.95 & --0.30 \\
      & (1010) & 5845.53 &         &        &
$F_2$ & (0003) & 3871.29 & 3870.49 &   0.80 \\
      & (0020) & 6003.65 & 6004.65 & --1.00 &
      & (0003) & 3931.36 & 3930.92 &   0.44 \\
      &        &         &         &        &
      & (0102) & 4143.09 & 4142.86 &   0.23 \\
$A_1$ & (0003) & 3909.20 & 3909.18 &   0.02 &
      & (0201) & 4349.01 & 4348.77 &   0.24 \\
      & (0102) & 4131.92 & 4132.99 & --1.07 &
      & (0201) & 4378.38 & 4379.10 & --0.72 \\
      & (0300) & 4595.26 & 4595.55 & --0.29 &
      & (1002) & 5523.80 &         &        \\
      & (1002) & 5498.66 &         &        &
      & (0012) & 5594.92 & 5597.14 & --2.22 \\
      & (0012) & 5617.16 &         &        &
      & (0012) & 5620.68 &         &        \\
      & (0111) & 5836.11 &         &        &
      & (0012) & 5632.36 &         &        \\
      & (1200) & 5973.26 &         &        &
      & (1101) & 5740.86 &         &        \\
      & (1011) & 7147.56 &         &        &
      & (0111) & 5830.28 &         &        \\
      & (0021) & 7300.85 &         &        &
      & (0111) & 5848.46 &         &        \\
      & (0120) & 7562.91 &         &        &
      & (0210) & 6054.58 &         &        \\
      & (3000) & 8583.81 &         &        &
      & (0210) & 6067.03 &         &        \\
      & (1020) & 8727.97 &         &        &
      & (2001) & 7094.16 &         &        \\
      & (0030) & 8975.64 & 8975.34 &   0.30 &
      & (1011) & 7145.84 &         &        \\
$A_2$ & (0102) & 4161.52 & 4161.87 & --0.35 &
      & (0021) & 7266.11 &         &        \\
      & (0300) & 4595.28 & 4595.32 & --0.04 &
      & (0021) & 7303.38 &         &        \\
      & (0111) & 5844.61 &         &        &
      & (0021) & 7344.87 &         &        \\
      & (0120) & 7550.53 &         &        &
      & (1110) & 7365.83 &         &        \\
$E$   & (0102) & 4105.22 & 4105.15 &   0.07 &
      & (0120) & 7514.67 &         &       \\
      & (0102) & 4152.15 & 4151.22 &   0.93 &
      & (2010) & 8594.90 &         &       \\
      & (0300) & 4592.13 & 4592.03 &   0.10 &
      & (1020) & 8786.05 &         &       \\
      & (1002) & 5535.04 &         &        &
      & (0030) & 8907.91 & 8906.78 &   1.13 \\
      & (0012) & 5620.36 &         &        &
      & (0030) & 9045.36 & 9045.92 & --0.56 \\
      & (0111) & 5836.45 &         &        &
      &        &         &         &        \\
& & & & & & & & & \\
\hline
\end{tabular}
\normalsize
\end{table}

\end{document}